\documentclass{article}

\usepackage{amsmath,amssymb,amsthm}

\newtheorem{theorem}{Theorem}
\newtheorem{mcor}{Corollary}
\newtheorem{mlem}{Lemma}

\theoremstyle{definition}

\newtheorem*{agradecimientos}{Acknowledgments}

\newcommand{\fref}[1]{\textnormal{(\ref{#1})}}

\begin{document}

\title{On the relation between the complex Toda and Volterra
lattices}

\author{Dolores Barrios Rolan\'\i a\\Universidad~Polit\'ecnica de Madrid (UPM)\\
Fac. Inform\'atica,
28660 Boadilla del Monte, Spain\\
E-mail: dbarrios@fi.upm.es\\[10pt]
 Rafael Hern\'andez Heredero\\ Universidad~Polit\'ecnica de
Madrid (UPM)\\
EUIT de Telecomunicaci\'on, 28031 Madrid, Spain \\
E-mail: rafahh@euitt.upm.es }

\maketitle

\begin{abstract}
\noindent
We give an analytic, sufficient condition for the existence of the B\"acklund
transformation between the semiinfinite Toda and Volterra lattices, in the
complex case, extending previous results given for the real case.
\end{abstract}

\section{Introduction}

The semiinfinite Toda lattice
\begin{equation}\label{Toda}
\left.
\begin{aligned}
\dot{\alpha}_n(t) & = \lambda_{n+1}^2(t)-\lambda_n^2(t)  \\
\dot{\lambda}_{n+1} (t)& =
\frac{\lambda_{n+1}(t)}{2}\left(\alpha_{n+1}(t)-\alpha_{n}(t)\right)
\end{aligned}\right\},\quad
n\in \mathbf{N}, \quad(\lambda_1\equiv0),
\end{equation}
(where dot means differentiation with respect to $t\in \mathbf{R}$) is a
well-known differential system with a remarkable property:
integrability~(cf.~\cite{Flaschka}). Integrability reveals itself as a multitude
of properties, like the existence of an infinite set of conservation laws,
symmetries, solutions that can be written explicitly, inverse scattering
solvability, etc. In this letter we center our attention in one of these
features: the existence of a so-called B\"acklund transformation that, from a
given solution of the system, produces another, different solution.

The case of the~\emph{real} Toda lattice, i.e. when the dependent
variables~$\alpha_{n}$ and~$\lambda_{n}$ are real functions,
 has been
thouroughly studied, being one of the cornerstones of modern
integrability theory. The~\emph{complex} case, when~$\alpha_{n}$
and~$\lambda_{n}$ are complex functions, has been found (much later) to have physical
applications, like the description of the asymptotic behaviour of some
N-soliton solutions of the nonlinear Schr\"odinger equation~\cite{Gerd}.
The algebraic nature of both
problems is the same, in the sense that they admit the same Lax pair
representation:
$$\dot{J}(t)=[J(t),K(t)]$$
is equivalent to~\fref{Toda}, where~$[J(t),K(t)]$ is the
commutator~$J(t)K(t)-K(t)J(t)$ of the operators represented by the semiinfinite
matrices
\begin{equation}
J(t)=\begin{pmatrix} \alpha_1(t) &  \lambda_2(t)  \\
 \lambda_2(t) & \alpha_2(t)& \lambda_3(t) \\
& \lambda_3(t) & \alpha_3(t)&\ddots &\\
&&\ddots&\ddots&
\end{pmatrix},\qquad K(t)=\frac 1 2 \begin{pmatrix} \phantom{-}0 &
-\lambda_2(t)
\\
 \lambda_2(t)  & 0& -\lambda_3(t)  \\
& \lambda_3(t)  &0&\ddots\\
&&\ddots&\ddots
\end{pmatrix}.
\end{equation}
Nevertheless, questions like the prolongability of solutions in time are not
easily treated by the techniques used in the real case. In fact, the first
reference to these problems we know of is the recent preprint~\cite{Gesz},
where an algebraic treatment is used.

As it has been said above, we are going to study here an special
property of the Toda lattice, the existence of a B\"acklund
transformation. For the real Toda lattice, this problem has been
analised in~\cite{Simon} and~\cite{Peherstorfer}. We extend this
analysis to the complex Toda lattice. We are going to provide an
analytical criterion, related with the spectral structure of the
matrix~$J(t)$. We believe it is the first criterion independent of
special algebraic structures, for the complex lattice case, and
one of its good features is that, once the correct setting and
structures have been introduced (cf.~\fref{polinomios}), the
mathematical proofs are straightforward, basically induction.

\section{The existence of the complex B\"acklund transformation}

The B\"acklund transformation appears through the intervention of a second
system, the semiinfinite Volterra or Langmuir lattice
\begin{equation}
\label{Langmuir}
\dot{\gamma}_{n+1}(t)=\gamma_{n+1}(t)\left(\gamma_{n+2}(t)-\gamma_n(t)\right),\quad
n\in \mathbf{N},\quad (\gamma_1\equiv 0).
\end{equation}
We call solution of system~\fref{Toda} (resp.~\fref{Langmuir}) to any sequence of differentiable, complex valued
functions of a real variable~$\{\lambda_n(t)$, $\alpha_n(t)\}$ (resp.~$\{\gamma_n(t)\}$), $n\in\mathbf{N}$,
satisfying the system.

Consider the tridiagonal matrix~$J(t)$. We use the representation
\begin{equation}
J(t)=\mathfrak{Re}J(t)+i\,\mathfrak{Im}J(t), \label{*}
\end{equation}
where, if~$A_{r,s}$ is the element in the~$r$-th row and $s$-th
column of a matrix~$A$, we
have~$(\mathfrak{Re}J(t))_{r,s}=\mathfrak{Re}J_{r,s}(t)$,
$(\mathfrak{Im}J(t))_{r,s}=\mathfrak{Im}J_{r,s}(t)$
and~$\mathfrak{Re} z $, $\mathfrak{Im} z $ denote the real and
imaginary part of~$z$. An important set of conditions will be
\begin{equation}\label{condiciones}
\mathfrak{Re}J(t) \mbox{ selfadjoint}, \quad
\left\|\mathfrak{Im}J(t)\right\| <+\infty,\quad t\in\mathbf{R},
\end{equation}
where $\left\|\mathfrak{Im}J(t)\right\|=\displaystyle\sup_{
\|x\|=1}\!\! \left\|\mathfrak{Im}J(t)x\right\| \,.$

\begin{theorem}\label{teorema1}
Let~$\{\lambda_n(t)\}$, $\{\alpha_n(t)\}$, $\{\gamma_n(t)\}$, $n\in \mathbf{N}$,
such that~\fref{condiciones} are satisfied and the relations
\begin{equation} \label{2}
\begin{aligned}
&\lambda_{1}(t)\equiv\gamma_{1}(t)\equiv0,\\
&\lambda_{n+1}^2(t)=\gamma_{2n}(t)\gamma_{2n+1}(t), \quad
\alpha_{n}(t)=\gamma_{2n-1}(t)+\gamma_{2n}(t)+C,\quad
n\in\mathbf{N}\,,
\end{aligned}
\end{equation}
hold for some~$C\in\mathbf{C}$ with
\begin{equation}\label{C}
d(C,\operatorname{Conv}\left(\sigma(\mathfrak{Re}J(t)))\right)
>\|\mathfrak{Im}J(t)\|\quad
\text{for any~$t\in \mathbf{R}$,}
\end{equation}
where~$\operatorname{Conv}(\cdot)$ denotes the convex hull and~$\sigma(\cdot)$
the spectrum.  Then~$\{\gamma_n(t)\}$, $n\in
\mathbf{N}$, is a solution of the Volterra lattice~\fref{Langmuir} if and only
if~$\{\lambda_n(t),\,\alpha_n(t)\}$, $n\in
\mathbf{N}$, is a solution of the Toda lattice~\fref{Toda}.
\end{theorem}
If~$J(t)$ is selfadjoint, then $\mathfrak{Im}J(t)=0$ in (\ref{*}).
In this case the constant~$C$ can be substituted by any value not
belonging to the convex hull of the spectrum of~$J(t)$, giving rise
to the following relation between the Volterra and real Toda
lattices.
\begin{mcor}\label{corolario1}
Let~$J(t)$ be selfadjoint for all~$t\in\mathbf{R}$. Consider
sequences~$\{\lambda_n(t)\}$, $\{\alpha_n(t)\}$, $\{\gamma_n(t)\}$, $n\in
\mathbf{N}$ satisfying~\fref{2} with~$C\notin
 \operatorname{Conv}\left(\sigma( J(t) )\right)$ for each~$t\in \mathbf{R}$.
Then~$\{\gamma_n(t)\}$, $n\in \mathbf{N}$, is a solution of~\fref{Langmuir} if
and only if~$\{\lambda_n(t),\,\alpha_n(t)\}$, $n\in\mathbf{N}$, is a solution
of~\fref{Toda}.
\end{mcor}
In addition, the relations~\fref{2} between solutions of~\fref{Toda}
and~\fref{Langmuir} yield a relation between two different solutions
of~\fref{Toda}, as stated in the following result, supplementary to
Theorem~\ref{teorema1}.
\begin{theorem}\label{teorema2}
Let~$\{\lambda_n(t),\,\alpha_n(t)\}$, $n\in\mathbf{N}$, be a solution
of~\fref{Toda} satisfying~\fref{condiciones}, and let~$C\in\mathbf{C}$ such
that~\fref{C} holds. Then there exists a
solution~$\{\widetilde\lambda_n(t),\,\widetilde\alpha_n(t)\}$, $n\in\mathbf{N}$,
of~\fref{Toda} and a solution~$\{\gamma_n(t)\}$, $n\in\mathbf{N}$
of~\fref{Langmuir}, such that
\begin{align}
\lambda_{n+1}^2(t)&=\gamma_{2n}(t)\gamma_{2n+1}(t), &
\alpha_{n}(t)&=\gamma_{2n-1}(t)+\gamma_{2n}(t)+C,& n\in\mathbf{N},\label{primera}\\
\widetilde\lambda_{n+1}^2(t)&=\gamma_{2n+1}(t)\gamma_{2n+2}(t), &
\widetilde\alpha_{n}(t)&=\gamma_{2n}(t)+\gamma_{2n+1}(t)+C,& n\in\mathbf{N}. \label{segunda}
\end{align}
For each fixed~$C\in \mathbf{C}$ satisfying~\fref{C}, the sequences~$\{\widetilde\lambda_n(t),\,
\widetilde\alpha_n(t)\}$, $\{\gamma_n(t)\}$, $n\in\mathbf{N}$, verifying~\fref{primera}, \fref{segunda} with~$
 \gamma_{1}(t)\equiv 0 $ are unique.
\end{theorem}

\section{Proofs}

Concerning Theorem~\ref{teorema1}, if~$\{\gamma_n(t)\}$ is a solution
of~\fref{Langmuir} and~$\{\lambda_n(t),\,\alpha_n(t)\}$ is taken as in~\fref{2}
with any~$C\in\mathbf{C}$, then this sequence of functions is immediately a
solution of~\fref{Toda}. Consequently, it is enough to prove
Theorem~\ref{teorema2} to complete the proof of Theorem~\ref{teorema1}. This is
the purpose of this section.

We consider that~$\{\lambda_n(t),\,\alpha_n(t)\}$ is a solution of~\fref{Toda}
such that~\fref{condiciones} are satisfied. The use of orthogonal polynomials
was proposed in~\cite{Peherstorfer} as a tool for proving Theorem~\ref{teorema2}
in the case when the operators~$J(t)$, $t\in \mathbf{R}$, are selfadjoint. Our
results are an extension to complex Toda lattices. We use the sequence of
polynomials (in~$z$) given by the recurrence
\begin{equation}\label{polinomios}
\left.
\begin{aligned}
&P_{n+1}(t,z)  =
(z-\alpha_{n+1}(t))P_{n}(t,z)-\lambda_{n+1}^2(t)P_{n-1}(t,z),\quad n\geq 0\\[2mm]
&P_{-1}(t,z)\equiv 0,\quad P_{0}(t,z)\equiv 1
\end{aligned}\,\right\}
\end{equation}
for each~$t\in\mathbf{R}$, which has the role of a parameter. The dependence
of~$\{P_n(t,z)\}$ on~$t$ is instrumental in order to establish the
relations~\fref{primera} and~\fref{segunda}. The following auxiliary result,
consequence of~\cite[Lemma 2, p.~523]{Peherstorfer}, and whose proof is
immediate by induction using~\fref{polinomios}, describes such dependence.
\begin{mlem}\label{lema1}
If~$\{\lambda_n(t),\,\alpha_n(t)\}$ is a solution of~\fref{Toda}, then
$$\dot{P}_n(t,z)=-\lambda_{n+1}^2(t)P_{n-1}(t,z),\quad n\in
\mathbf{N},$$
where the derivative~$\dot{P}_n(t,z)$ is taken with respect to~$t\in\mathbf{R}$.
\end{mlem}
For each~$t\in\mathbf{R}$ the set of zeros of each polynomial~$P_n(t,z)$ is the spectrum of
$J_n(t)\,,\,\sigma\left(J_n(t)\right)$, being~$J_n(t)$ the $n\times n$ submatrix formed by the first $n$ rows
and columns of~$J(t)$. The relation between the spectra of~$J(t)$ and its real part was studied
in~\cite{finite}, and we need the following result, derived there. We suppose that~$C$ satisfies~\fref{C}. The
expressions~$\mathfrak{Re} J_n(t)$ and~$\mathfrak{Im} J_n(t)$ denote the submatrices formed by the first $n$
rows and columns of~$\mathfrak{Re} J(t)$ and~$\mathfrak{Im} J(t)$, respectively, and~$\|\cdot\|$ is the norm of
each operator in the space where it is defined, i.e.~either~$\ell^2$ or~$\mathbf{C}^n$.
\begin{mlem}\label{lema2} $(${\normalfont cf. \cite[\text{lemmas 1 and 2}]{finite} }$)$
With the restrictions~\fref{condiciones}, for each~$n\in\mathbf{N}$ we have
\begin{itemize}
\item[1.-] $\|\mathfrak{Im} J_n(t)\|\leq\|\mathfrak{Im} J(t)\|$
\item[2.-] If $d\left(C,\sigma(\mathfrak{Re}J_n(t))\right)>
\|\mathfrak{Im}J_n(t)\|$ then $P_n(t,C)\neq 0$
\end{itemize}
\end{mlem}
For~$n\in\mathbf{N}$ fixed, from the well-known fact
that~$\sigma\left(\mathfrak{Re}J_n(t)\right)\subset
\operatorname{Conv}\left(\sigma(\mathfrak{Re}J(t))\right)$ and Lemma~\ref{lema2}
it follows that~$P_n(t,C)\neq 0$.

As in~\cite[Th.~1]{Peherstorfer}, consider the sequence of monic
polynomials~$\{Q^{(C)}_n(t,z)\}$ defined by
\begin{equation}
Q^{(C)}_n(t,z)= \frac{P_{n+1}(t,z) - \frac{P_{n+1}(t,C)}{P_{n}(t,C)}P_n(t,z)}{z-C},\quad n=0,1,\ldots , \quad
t\in \mathbf{R} \label{7}
\end{equation}
(cf.~\cite[p.~35]{Chihara}). Let us prove that these polynomials satisfy a three-term recurrence relation. The
fact, though, that the functions~$ \lambda_n(t),\,\alpha_n(t) $, $n\in\mathbf{N}$, are complex-valued, prevents
the use of the standard proof using the corresponding moment functional, because now it is not
positive-definite. Nevertheless, a direct proof is still possible.
\begin{mlem}\label{lema3}
The sequence of polynomials~$\{Q^{(C)}_n(t,z)\}$ satisfies the three-term
recurrence relation
\begin{equation}\left.\begin{aligned}
&Q^{(C)}_{n}(t,z)  =
(z-\widetilde\alpha_{n}(t))Q^{(C)}_{n-1}(t,z)-\widetilde\lambda_{n}^2(t)Q^{(C)}_{n-2}(t,z),\quad
n\in\mathbf{N}\\[2mm]
&Q^{(C)}_{-1}\equiv 0,\quad Q^{(C)}_{0}\equiv 1
\end{aligned}\,\right\},\label{8}\end{equation}
being
\begin{equation}\left.
\begin{aligned}
\widetilde\alpha_{n}(t)& = \frac{P_{n+1}(t,C)}{P_{n}(t,C)}+\alpha_{n+1}(t)-  \frac{P_{n}(t,C)}{P_{n-1}(t,C)}
\\
\widetilde\lambda_{n}^2(t)& = \lambda_n^2(t)\frac{P_{n-2}(t,C)P_{n}(t,C)}{P^2_{n-1}(t,C)}
\end{aligned}
\right\},\quad n\in\mathbf{N},\quad t\in\mathbf{R}.\label{9}
\end{equation}
\end{mlem}
\noindent\underline{Proof of Lemma~\ref{lema3}}. Formula~\fref{7}
gives~$Q^{(C)}_{0}(t,z)\equiv 1$, and also gives the same expression
than~\fref{8} for~$Q^{(C)}_{1}(t,z)$. The sequence of polynomials given
by~\fref{8} is unique. Thus, it is enough to check that, for~$n\geq2$, the
sequence of polynomials given by~\fref{7} satisfy~\fref{8}.
Substituting~$(z-C)Q^{(C)}_{n-1}(t,z)$ and~$(z-C)Q^{(C)}_{n-2}(t,z)$ given
by~\fref{7} in the right hand side~of~\fref{8} yields (suppressing explicit
$t$-dependence for brevity)
\begin{multline}\label{10}
(z-\widetilde \alpha_n)\left(P_n(z)-\frac{P_n(C)}{ P_{n-1}(C)} P_{n-1}(z)\right) - \widetilde\lambda_n^2
\left(P_{n-1}(z)-\frac{P_{n-1}(C)}{ P_{n-2}(C)} P_{n-2}(z)\right)=
\\
(z-\widetilde \alpha_n )P_n(z)-\left[(z-\widetilde \alpha_n)\frac{P_n(C)}{ P_{n-1}(C)}+\widetilde \lambda_n^2
\right]P_{n-1}(z) +\widetilde \lambda_n^2 \frac{P_{n-1}(C)}{ P_{n-2}(C)}P_{n-2}(z).
\end{multline}
From~\fref{9} we have $ \widetilde \lambda_n^2\frac{P_{n-1}(C)}{P_{n-2}(C)} =\lambda_n^2
\frac{P_{n}(C)}{P_{n-1}(C)}$, and taking into account also~\fref{polinomios}, $ z-\widetilde \alpha_n
=z-C+\lambda_{n+1}^2\frac{P_{n-1}(C)}{P_{n}(C)}+ \frac{P_{n}(C)}{P_{n-1}(C)}$, $n\in\mathbf{N}$. Thus~\fref{10}
becomes
\begin{multline*}
\left(z-\alpha_{n+1} -\frac{P_{n+1}(C)}{ P_{n}(C)} +\frac{P_{n}(C)}{ P_{n-1}(C)}\right)P_n(z)- \left(
(z-C)\frac{P_{n}(C)}{ P_{n-1}(C)} +\lambda_{n+1}^2 \right.
\\
\left.{}+\frac{P^2_{n}(C)}{P^{2}_{n-1}(C)} +\lambda_{n}^2 \frac{P_{n-2}(C)P_{n}(C)}{
P^{2}_{n-1}(C)}\right)P_{n-1}(z) +\lambda_{n} \frac{P_{n}(C)}{ P_{n-1}(C)}P_{n-2}(z) \\=(z-\alpha_{n+1} )P_n(z)-
\lambda_{n+1}^2 P_{n-1}(z)-\frac{P_{n+1}(C)}{P_{n}(C)}P_n(z)+ \frac{P_{n}(C)}{P_{n-1}(C)}P_n(z)
\\
-\frac{P_{n}(C)}{P_{n-1}(C)}\left(z-C+ \frac{P_{n}(C)}{P_{n-1}(C)}+\lambda_n^2
\frac{P_{n-2}(C)}{P_{n-1}(C)}\right)P_{n-1}(z) +\lambda_n^2 \frac{P_{n}(C)}{P_{n-1}(C)}P_{n-2}(z)
\end{multline*}
Using~\fref{polinomios} in the last expression, we get
\begin{multline*}
P_{n+1}(z)
 -\frac{P_{n+1}(C)}{P_{n}(C)}P_n(z)+
\frac{P_{n}(C)}{P_{n-1}(C)}P_n(z)-\frac{P_{n}(C)}{P_{n-1}(C)}(z-\alpha_n )P_{n-1}(z)\\
+\lambda_n^2
\frac{P_{n}(C)}{P_{n-1}(C)}P_{n-2}(z)=P_{n+1}(z)-\frac{P_{n}(C)}{P_{n-1}(C)}P_n(z),
\end{multline*}
which is $(z-C)Q^{(C)}_n(z)$. \hfill $\square$

Lemma~\ref{lema1} and~\fref{9} imply that the
sequence~$\{\widetilde\lambda_n(t),\,\widetilde\alpha_n(t)\}$, $n\in\mathbf{N}$,
is a solution of the Toda lattice~\fref{Toda}. Define the
sequence~$\{\gamma_n(t)\}$, $n\in\mathbf{N}$, as
\begin{equation}\gamma_{1}(t)=0,\quad
\gamma_{2n}(t)=-\frac{P_{n}(t,C)}{P_{n-1}(t,C)},\quad
\gamma_{2n+1}(t)=-\lambda_{n+1}^2(t) \frac{P_{n-1} (t,C)}{
P_{n}(t,C)}. \label{catorce}
\end{equation}
Lemma~\ref{lema1} again implies that~$\{\gamma_n(t)\}$ is a solution of the
Volterra lattice~\fref{Langmuir}. Besides, the relations~\fref{primera}
and~\fref{segunda} are readily checked.

Uniqueness of the sequences obtained satisfying~\fref{primera} and~\fref{segunda} can be checked directly.
Suppose that there exists another sequence~$\{\widetilde \gamma_n(t)\}$ with~$\tilde\gamma_{1}(t)\equiv 0$.
Then~\fref{primera} and~\fref{segunda} imply that
\begin{gather}
\gamma_{2n-1}(t)+\gamma_{2n}(t)=\widetilde\gamma_{2n-1}(t)+\widetilde\gamma_{2n}(t),\quad
n\in \mathbf{N}, \label{I}\\
\gamma_{2n}(t)\gamma_{2n+1}(t)=\widetilde\gamma_{2n}(t)\widetilde\gamma_{2n+1}(t),\quad
n\in \mathbf{N}. \label{II}
\end{gather}
For~$n=1$, \fref{I} means that~$\gamma_{2}(t)=\widetilde \gamma_{2}(t)$ and then~\fref{II} leads
to~$\gamma_{3}(t)=\widetilde \gamma_{3}(t)$. In general, if~$\gamma_{2n-1}(t)=\widetilde \gamma_{2n-1}(t)$,
\fref{I} implies that~$\gamma_{2n}(t)=\widetilde \gamma_{2n}(t)$ and~\fref{II} that~$\gamma_{2n+1}(t)=\widetilde
\gamma_{2n+1}(t)$ (note that, because of (\ref{catorce}), we have~$\gamma_{2m}(t)\neq0$, $m\in \mathbf{N}$). So,
both sequences coincide. The uniqueness of~$\{\widetilde\lambda_n(t),\, \widetilde\alpha_n(t)\}$ follows from
the uniqueness of~$\{\gamma_n(t)\}$ and from~\fref{segunda}. This proves Theorem~\ref{teorema2}.

\section{Remarks and Conclusions}

Theorem~\ref{teorema2} provides a method for constructing families
of solutions of~\fref{Toda} from a given solution, real or complex.
Choosing~$C\notin \mathbf{R}$ provides new complex solutions
of~\fref{Toda} and~\fref{Langmuir}. On the contrary, if~$C\in
\mathbf{R}\setminus \operatorname{Conv}(\sigma(J(t)))$ for all~$t\in
\mathbf{R}$,  and the starting solution is real, the
relations~\fref{9} and~\fref{catorce} show that the new solutions
are also real, with~$\widetilde \lambda_n(t)$, $\gamma_n(t)>0$, $t\in
\mathbf{R}$, $n\in \mathbf{N}$.

On the other hand, in theorems~\ref{teorema1} and~\ref{teorema2}
some restriction on the value of~$C$ is necessary. We can understand
this fact by taking arbitrary sequences~$\{\lambda_{n}(t)\}$,
$\{\alpha_{n}(t)\}$, $\{\gamma_{n}(t)\}$ with~$\lambda_{n}(t)\neq0$,
$n\geq2$, $t\in\mathbf{R}$ $($not necessarily solutions
of~\fref{Toda} and~\fref{Langmuir}$)$, and~$C\in\mathbf{C}$
verifying~\fref{2}. Then, if~$\{P_n(t,z)\}$ is the sequence of
polynomials given by~\fref{polinomios} we have $P_n(t,C)\neq 0
\mbox{ for all $t\in \mathbf{R}$.}$ In fact, these conditions imply
that the~$\{\gamma_n(t)\}$ given in~\fref{catorce} is the only
sequence verifying~\fref{2}, as can be deduced from the proof of
Theorem~\ref{teorema2}.

With the premises and the notation given in Theorem~\ref{teorema2}
define, for each~$t\in\mathbf{R}$, the polynomials
$$S_{2n}^{(C)}(t,z)=P_n(t,z^2+C) ,\quad
S_{2n+1}^{(C)}(t,z)=zQ_n(t,z^2+C),\quad n\in\mathbf{N}\,.$$ Using
(\ref{polinomios}) and (\ref{7}) it is easy to see that the sequence
$\{S_{n}^{(C)}(t,z)\},\,n\in \mathbf{N},$ verify the recurrence
relation
$$
\left.
\begin{aligned}
&S^{(C)}_{n}(t,z)  =
zS^{(C)}_{n-1}(t,z)-\gamma_{n}(t)S^{(C)}_{n-2}(t,z),\quad n\in
\mathbf{N} \\
&S^{(C)}_{-1}\equiv 0,\quad S^{(C)}_0\equiv 1
\end{aligned}\,\right\}.$$
In other words, the relations between the sequences~$\{P_n\}$,
$\{Q_n^{(C)}\}$, $\{S_n^{(C)}\}$ are analogous to those given
in~{\normalfont\cite[Th.~9.1, p.~46]{Chihara}}, corresponding to
polynomials independent of~$t$ associated with constants that
satisfy~\fref{primera} and~\fref{segunda}.

\begin{agradecimientos}This work was partially supported by
Universidad Polit\'{e}cnica de Madrid, under grant AY05/11263. The
research of D. Barrios Rolan\'{\i}a was partially supported by
Direcci\'{o}n General de Investigaci\'{o}n, Ministerio de Educaci\'{o}n y
Ciencia, under grant BFM 2003-06335-C03-02
\end{agradecimientos}

\end{document}